\documentclass[conference]{IEEEtran}
\ifCLASSINFOpdf

\else

\fi

\usepackage[caption=false,font=footnotesize]{subfig}
\usepackage{graphicx}
\usepackage{soul}
\usepackage[linesnumbered,ruled]{algorithm2e}
\usepackage{subfig}
\usepackage[fleqn]{amsmath}
\usepackage{relsize}
\usepackage{amsmath}
\let\labelindent\relax
\usepackage{enumitem}

\begin{document}
%
\title{LBICA: A \underline{L}oad \underline{B}alancer for \underline{I}/O \underline{C}ache \underline{A}rchitectures\vspace{-1em}}

%

\author{\IEEEauthorblockN{Saba Ahmadian, Reza Salkhordeh, and Hossein Asadi}
	\IEEEauthorblockA{Data Storage, Networks, and Processing (DSN) Lab, Department of Computer Engineering\\
		Sharif University of Technology, Tehran, Iran
		\\ Email: \{ahmadian, salkhordeh\}@ce.sharif.edu, and asadi@sharif.edu}
}

\maketitle

\begin{abstract}

In recent years, enterprise \emph{Solid-State Drives} (SSDs) are used in the caching layer of high-performance servers to close the growing performance gap between processing units and storage subsystem.
SSD-based I/O caching is typically not  effective in workloads with burst accesses in which the caching layer itself becomes the performance bottleneck because of the large number of accesses.
Existing I/O cache architectures mainly focus on maximizing the cache hit ratio while they neglect the average queue time of accesses.
Previous studies suggested bypassing the cache when burst accesses are identified.
These schemes, however, are \emph{not} applicable to a general cache configuration and also result in significant performance degradation on burst accesses.

In this paper, we propose a novel I/O cache load balancing scheme (LBICA) with adaptive write policy management to prevent the I/O cache from becoming performance bottleneck in burst accesses.
Our proposal, unlike previous schemes, which disable the I/O cache or bypass the requests into the disk subsystem in burst accesses, selectively reduces the number of waiting accesses in the SSD queue and balances the load between the I/O cache and the disk subsystem while providing the maximum performance.
The proposed scheme characterizes the workload based on the type of in-queue requests and assigns an effective cache write policy.
We aim to bypass the accesses which 1) are served faster by the disk subsystem or 2) cannot be merged with other accesses in the I/O cache queue.
Doing so, the selected requests are responded by the disk layer, preventing from overloading the I/O cache.
Our evaluations on a physical system shows that LBICA reduces the load on the I/O cache by 48\% and improves the performance of burst workloads by 30\% compared to the latest state-of-the-art load balancing scheme.

\end{abstract}


%
\IEEEpeerreviewmaketitle
\section{Introduction}
\label{sec:introduction}
\begin{figure}[!h]
	\centering
	\includegraphics[scale=1.2]{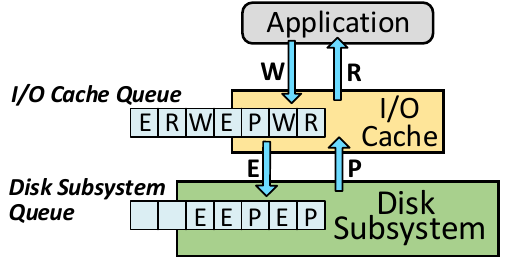}
		\vspace{-0.2em}
	\caption{I/O cache and disk subsystem queue in a write-back cache (R: Read, W: Write, P: Promote, and E: Evict).}
		\vspace{-0.5em}
	\label{fig:io_load}
\end{figure}

Increasing number of I/O intensive applications such as \emph{Online Transaction Processing} (OLTP), \emph{High Performance Computing} (HPC), web, and email applications arises the demand in data-centers for high-performance storage systems. The most common approach to improving the performance of storage systems is to employ \emph{Solid-State Drives} (SSDs) \cite{tarihi2016hybrid} in the caching layer of the disk subsystems \cite{ahmadian2018eci,reca,tica,ahmadian-ssd-rel-date,salkhordeh2015operating}, which are mainly built upon low-performance and low-reliable \emph{Hard Disk Drives} (HDD) \cite{kishani-tr-2018,kishani2017evaluating,kishani-tr-2018-2} or mid-range SSDs (as shown in Fig. \ref{fig:io_load}).
Inclusion of  SSDs in the I/O caching layer of systems improves the response time of the requests supplied by the cache, and hence, a wide range of enterprise and academic I/O cache architectures are proposed with the purpose of maximizing the hit ratio of the caching layer \cite{emc_fast_cache-online,hp_smart_cache-online,netapp_ssd_cache-online,ahmadian2018eci,hystor,reca, vcacheshare, scave, centaur,janus,mercury, migratetossd,xia2016high,matthews2008intel,kim2011hybridstore,nitro,azor,xia2015flash, jin2017kaml, shen2017didacache, cao2018alacc}.  
In these I/O caching schemes, mainly based on \emph{datapath} or \emph{push mode} cache architectures, the entire accesses are directed  to the caching layer \cite{wec2015} and as such, the highest number of requests is responded  via the caching layer to  achieve the highest performance in terms of hit ratio. 
In addition, \emph{non-datapath} caching schemes \cite{wec2015} have been proposed, which mainly focus on improving the endurance of the SSD cache with considerable performance overhead.

To obtain the highest performance for both read and write accesses, storage architects offer assigning \emph{Write Back} (WB) policy on the caching layer. Doing so, both read and write accesses are buffered in the cache, which results in improving the performance of future accesses to the same blocks.
However, recent studies suggested using cache write policies other than WB, such as \emph{Write Through} (WT), \emph{Read Only} (RO), and \emph{Write Only} (WO) to provide higher reliability or improve the endurance of the SSD cache while usually have performance overheads on the running workloads \cite{ahmadian2018eci}. 
Using WB policy (as enterprise approaches do \cite{emc_fast_cache-online,hp_smart_cache-online,netapp_ssd_cache-online}), in contrast, imposes unwanted I/O load on both the caching layer and the disk subsystem (as shown in Fig. \ref{fig:io_load}). For instance, each missed read request is supplied at the cost of imposing one read access on the disk subsystem and one write access to the caching layer (due to promoting new data block). This becomes worse when an eviction is required to provide free cache blocks before promoting the newer ones.

Existing caching architectures, mainly with the purpose of providing the highest hit ratio, cannot overcome the heavy I/O load of the burst accesses of applications such as boot storms, OLTP, TPC-C, mail, web, and database servers \cite{kim2018selective}.
The burst accesses of such heavy workloads remain in the queue of the caching layer (since they will be served by cache, i.e., cache hit) while the disk subsystem is idle without serving any request. Thus, the caching layer becomes the performance bottleneck of the system \cite{kim2018selective}.
Such poor load balancing between layers of storage hierarchy is due to 1) ignoring the queue time of the I/O requests in the caching layer and 2) not considering the impact of promotion and eviction of data blocks on the cache I/O load.
Recent studies suggest I/O bypassing schemes, in which they estimate the wait time of in-queue requests and selectively direct the requests with the highest estimated latencies to the disk subsystem \cite{kim2018selective}.  Such schemes suffer from three main shortcomings: 1) they \emph{only} consider a specific caching policy where they would not be applicable to the other types of I/O caches. In addition, such cache policy (WT and WO) is not employed in the caching layer of enterprise storage systems. This is because it only can improve the performance of read after write accesses while read and write requests experience the latency of disk subsystem.
2) The way how they select to-be-bypassed requests is inefficient, which imposes a considerable performance overhead on the operation of the queue.
3) Such schemes may bypass the requests, which were supposed to be hit in the cache while they may keep the requests in the cache queue which would not improve the hit ratio resulting in a significant performance degradation in burst accesses.

In this paper, we propose LBICA, a novel I/O cache load balancing scheme, which adaptively assigns an effective write policy to the caching layer. LBICA 1) prevents the caching layer from becoming the performance bottleneck and 2) improves the performance of the workloads with burst I/O accesses.
Unlike previous load balancing schemes, which are optimized to a specific cache policy, our proposal is applicable to different caching architectures. Furthermore, LBICA 1) detects burst workloads and 2) characterizes the requests of the workloads, then adaptively assigns effective write policies to the caching layer.
Doing so, we eliminate the performance overhead of selecting the to-be-bypassed requests and only bypass the accesses which 1) have not considerable impact on the overall performance and 2) cannot be merged with the other accesses in the caching layer.
LBICA, using kernel level tools such as \emph{iostat} \cite{iostat_online} and \emph{blktrace} \cite{blktrace}, detects the burst intervals and also characterizes the workloads and puts them in a) \emph{random read}, b) \emph{random write}, c) \emph{sequential read}, d) \emph{sequential write}, and e) \emph{mixed read and write} groups.  We assign efficient write policy to the caching layer based on the characteristics of the running workloads in different burst intervals, which results in enhanced performance and highly utilized storage subsystem.
We evaluate LBICA on a physical system including 4TB SAS 7.2K Seagate HDD in the disk subsystem level and 1TB 863a Samsung SSD in the I/O cache level. We use EnhanceIO \cite{enhanceio}, as an open-source I/O caching kernel module, to develop the proposed caching architecture. Our experimental results show that the proposed I/O cache load balancing scheme reduces the load on the I/O cache by 48\% and improves the performance by 30\% compared to the latest state-of-the-art load balancing scheme. 
We make the following contributions: 
\begin{enumerate}[leftmargin=*]
	\item We propose LBICA, a novel I/O cache load balancing scheme, which effectively improves the performance and prevents the bottleneck effect of I/O caching layer in burst accesses.
	\item LBICA considers the requests as a) application read, b) application write, c) cache promote, and d) cache evict, and characterizes the running workloads based on the in-queue requests type.
	\item Our proposed load balancing scheme effectively assigns efficient cache write policy in different burst intervals based on the I/O characteristics of the running workloads.
	\item Our proposal effectively bypasses the I/O requests from the cache while serving such request from the disk subsystem would not have any negative impact on the overall performance resulting in reduced I/O cache queue time.
\end{enumerate}

\section{Related Work}
\label{sec:related}
Existing SSD-based I/O caching schemes such as Janus \cite{janus}, Hystor \cite{hystor}, ReCA \cite{reca}, KAML \cite{jin2017kaml}, DIDACache \cite{shen2017didacache}, and ALACC \cite{cao2018alacc} mainly focus on  maximizing performance (in terms of hit ratio) of the running workloads. 
Such schemes aim to direct all I/O requests into the cache, and hence reduce the response time of the accesses, which are served by the caching layer. In addition, recent studies consider the impact of such I/O caching schemes on the lifetime of SSDs by proposing the I/O caching schemes to reduce the number of writes on the SSDs with the minimum performance overhead. 
Such I/O caching architectures do not provide the mechanisms to overcome the heavy load of burst I/O accesses due to OLTP or database applications. Thus, by neglecting the queue time of the requests in both I/O cache and disk queues, the caching layer behaves as a performance bottleneck resulting in a significantly large latency.
However, only few studies consider the impact of queue time of the requests in the performance provided by the I/O cache. Such studies aim to enhance the average I/O latency of the workloads by balancing the load on the I/O cache and disk subsystem. 
\emph{Selective I/O Bypass} (SIB) \cite{kim2018selective} is the latest state-of-the-art load balancing schemes, which aims to balance the load of I/O requests between SSD and disk subsystem preventing the I/O cache to become performance bottleneck. This scheme is designed in a way that \emph{only} works for WT‌ and WO caches in which only write accesses are buffered in the I/O cache while they are propagated to the disk subsystem at the same time. Such cache policy mainly aims to preserve the reliability with the following shortcomings which prevents the storage designers from employing such I/O cache scheme: 1) it only improves the performance of read accesses, which would be supplied by the cache (i.e., read after write accesses), 2) there is no any performance improvement on other accesses such as read after read (due to WO policy in which no read access is buffered in the cache) and write accesses (due to WT policy in which the accesses experience the latency of disk subsystem).

SIB selectively bypasses the I/O requests of the burst accesses into the disk subsystem. To do so, this scheme estimates the latency of in-queue requests and directs them to the disk subsystem \cite{kim2018selective}. 
The major disadvantages of SIB which we aim to resolve are: 1) employing the WT‌ policy on the SSD cache. In such WT‌ cache, since all write requests are supplied by both cache and disk in the same time, the queue size of both the cache and disk becomes the same in write-intensive workloads. In addition, due to higher delay of the disk compared to the SSD, the queue size of the disk becomes larger than that of SSD. In such condition, in case of burst accesses of a write-intensive workload, both SSD and disk become overloaded where no load balancing is possible. This is because the bypassed requests from the SSD cache will experience much larger delay in the disk queue.
2) selecting to-be-bypassed requests imposes performance and computational overhead on the system. 
3) Such I/O cache policy (WT and WO) is not applicable on enterprise systems due to negligible performance improvement which only affects read after write accesses.
In summary, SIB is the most close approach to LBICA with the above-mentioned shortcomings where our proposal aims to resolve them by 1) applying adaptive write policies on the I/O cache and 2) preventing unnecessary promotion (or eviction) to (or from) the cache which leads to eliminating significant unwanted I/O load on both SSD‌ cache and disk subsystem.

\section{Proposed Method}
\label{sec:proposed}
\begin{figure}[!h]
	\centering
		\vspace{-1.2em}
	\includegraphics[scale=1.4]{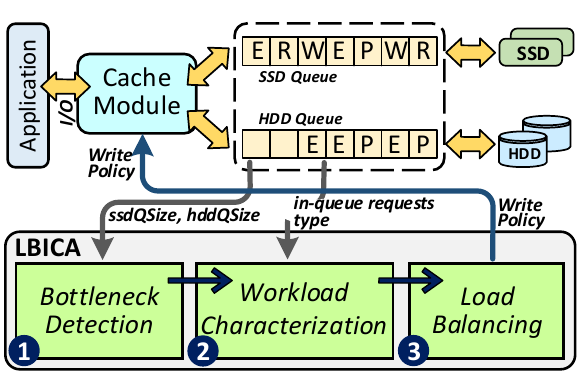}
		\vspace{-0.5em}
	\caption{The proposed I/O cache load balancing scheme (R: Read, W: Write, P: Promote, and E: Evict).}
		\vspace{-1em}
	\label{fig:LBICA}
\end{figure}

In this section, we propose our I/O cache load balancing scheme (LBICA).
As shown in Fig. \ref{fig:LBICA}, LBICA consists of three main procedures in which it 1) detects heavy I/O load, 2) characterizes the running workload, and 3) balances the I/O load between storage hierarchy.
LBICA gets the SSD and HDD queue size and based on the in-queue requests characteristics, it decides an efficient write policy for the cache to provide both load balance and I/O performance.
In Section \ref{sec:detection} and Section \ref{sec:char}, we show how LBICA detects burst I/O accesses and characterizes the workload, respectively. Then in Section \ref{sec:balance}, we elaborate the proposed load balancing scheme.


\subsection{Bottleneck Detection}
\label{sec:detection}
To detect the burst interval (i.e., heavy I/O load) on the storage, LBICA uses \emph{iostat} \cite{iostat_online} from \emph{sysstat} package as an comprehensive kernel level I/O tool which reports the I/O statistics in the block layer.
Periodically, using the given information by \emph{iostat}, we calculate the maximum queue time of both the I/O cache and disk subsystem based on Eq. \ref{eq:qtime}. 
\begin{equation}
\begin{aligned}
\label{eq:qtime}
cache~Qtime = ssdQSize \times ssdLatency\\\newline
disk~Qtime = hddQSize \times hddLatency
\end{aligned}
\end{equation}
Where $cache~Qtime$ and $disk~Qtime$ are the maximum queue time of the I/O cache and disk subsystem, respectively.
$ssdQSize$ and $hddQSize$ are the current queue size of the cache and disk subsystem (i.e., the number of pending requests in the queue), respectively. $ssdLatency$ and $hddLatency$ are the average read/write latency (i.e., service time) of the employed SSD‌ and HDD in the caching and disk subsystem level.
We detect the I/O caching layer as performance bottleneck when the maximum latency of the in-queue requests in I/O cache ($cache~Qtime$) is greater than that of the disk subsystem ($disk~Qtime$). In this case, bypassing the in-queue requests from the cache to the disk subsystem results in improved response time. In the following we elaborate on how we select and bypass the group of the requests to the disk subsystem.

\subsection{Workload Characterization}
\label{sec:char}

In case of burst I/O intervals where the I/O cache becomes the performance bottleneck, LBICA aims to characterize the running workload based on in-queue requests. We use \emph{blktrace} \cite{blktrace} as a block level I/O tracing tool to get the list of in-queue requests.
 The in-queue requests in the I/O cache can be either read  or write, where each of them may be due to 1) accesses of application (shown as R: Read and W: Write) or 2) accesses due to promotion and eviction of data blocks (shown as P and E). 
Based on the ratio of requests type (R: Read, W: Write, P: Promote, or E: Evict), we put the workloads in the following groups:\footnote{We assume that the workload has passed its warm-up interval.}

\begin{figure}[!h]
	\centering
	\subfloat[Random Read]{\includegraphics[width=.23\textwidth]{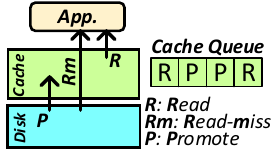}%
		\label{fig:rand-read}}
	\hfil
	\hspace{-0.4em}
	\subfloat[Mixed Read Write]{\includegraphics[width=.23\textwidth]{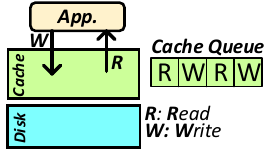}%
		\label{fig:mixed}}
	\hfil
	\subfloat[Write Intensive]{\includegraphics[width=.23\textwidth]{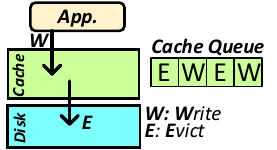}%
		\label{fig:write-intensive}}
	\hfil
	\hspace{-0.4em}
	\subfloat[Seq. Read]{\includegraphics[width=.23\textwidth]{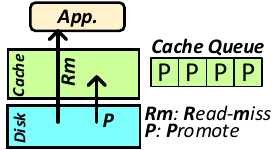}%
		\label{fig:seq-read}}
	
	\caption{Workload characterization based on in-queue requests type by LBICA.}
	\label{fig:work-char}
\end{figure}

\begin{enumerate}[leftmargin=*]
	\item \textbf{Group\_1}: mainly includes the requests from R and P types (shown in Fig. \ref{fig:rand-read}). Such accesses represent the behavior of  a workload with \emph{random read} access pattern in which most of the accesses are served by the cache (hit) and remaining are supplied by the disk subsystem (miss) and are promoted to the caching layer.
	
	\item \textbf{Group\_2}: mainly includes R and W types of the requests (shown in Fig. \ref{fig:mixed}). Such accesses are due to running a \emph{mixed read write} workload in the application level where the written data blocks in the I/O cache are accessed (read) by the future requests (i.e., hit).
	
	\item \textbf{Group\_3}: mainly includes W and E requests (shown in Fig. \ref{fig:write-intensive}), which shows the accesses of a \emph{write intensive} workload. In case of higher ratio of W compared to E, we detect this workload as \emph{random write}. Otherwise, the workload is categorized as \emph{sequential write}.
	
	\item \textbf{Group\_4}: mainly includes the requests from P type (shown in Fig. \ref{fig:seq-read}). Such accesses represent a \emph{sequential read} workload in which all read accesses are missed from cache and are promoted to the caching layer.
\end{enumerate}
The remaining groups are the set of accesses with 1) majority of R and E and 2) majority of W and P. Such set of accesses may not occur during a workload execution on a storage subsystem with I/O caching architecture, and hence, we do not consider them in our proposed workload characterization.

\subsection{Load Balancing}
\label{sec:balance}
In case of burst accesses, to balance the I/O load between the caching layer and disk subsystem, LBICA assigns an efficient write policy to the cache based on the characteristics of the running workload (as elaborated in Section \ref{sec:char}).
In the following, we first provide our proposed efficient cache write policy assignment and then show how such scheme 1) prevents the I/O cache bottleneck effect and 2) provides maximum performance (in terms of latency) for the running workload.
\begin{enumerate}[leftmargin=*]
	\item We set the \emph{Write Only} (WO) policy on the I/O cache when the running  workload is from Group\_1 category (i.e., random read workload).
	Such policy assignment provides the following benefits: 1) the I/O cache serves the read accesses (i.e., hit) and 2) read misses which are supplied by the disk subsystem are not promoted to the cache, reducing heavy load of writes due to promotions on the I/O cache. 
	
	\item When the running workload on the disk subsystem is categorized in Group\_2 (i.e., mixed read write workload), we set \emph{Read Only} (RO) policy on the cache. Doing so, we reduce the load on the I/O cache with bypassing the write accesses to the disk subsystem and only serve read accesses from the caching layer. The reason behind this policy is the higher priority of read accesses over writes.
	
	\item The write policy of the cache is set to WB in the workloads from Group\_3 and only the requests from the tail of the cache queue are bypassed to the disk subsystem. Such approach 1) provides the highest available performance for the requests of the cache queue which are below of the bottleneck threshold and 2) supplies the requests which are in the bottleneck threshold of the I/O cache by disk subsystem and provides smaller latency compared to the queue time of the I/O cache.  
	
	\item We set the WB policy on the cache when the running workload is from Group\_4. This is because the caching layer has no impact on the supplying of sequential read accesses, which are fully responded by the disk subsystem (i.e., cache miss). In this case, the I/O cache never becomes performance bottleneck. 
\end{enumerate}

\section{Experimental Results}
\label{sec:results}
In this section, we evaluate  LBICA  and show how our proposed I/O cache load balancing scheme provides higher performance compared to the previous load balancing schemes. We compare LBICA with 1) baseline WB cache without load balancing  and 2) SIB as the latest state-of-the-art load balancing scheme \cite{kim2018selective}. To do so, we develop both schemes and perform the same set of experiments in our physical test platform.

\subsection{Experimental Setup}
We evaluate LBICA on a physical system, where we use 4TB SAS 7.2K Seagate HDD in the disk subsystem and 1TB SSD Samsung 863a in the I/O caching layer.
We implement I/O cache level by an open-source, EnhanceIO \cite{enhanceio}, kernel level module. 
We run different types of workloads from TPC-C, mail server, and web server with burst I/O accesses and compare the I/O load and performance of different architectures.  

\subsection{I/O Load Comparison}
\begin{figure*}[!htb]
	\centering
		\vspace{-1em}
	\subfloat[TPC-C]{\includegraphics[width=.31\textwidth]{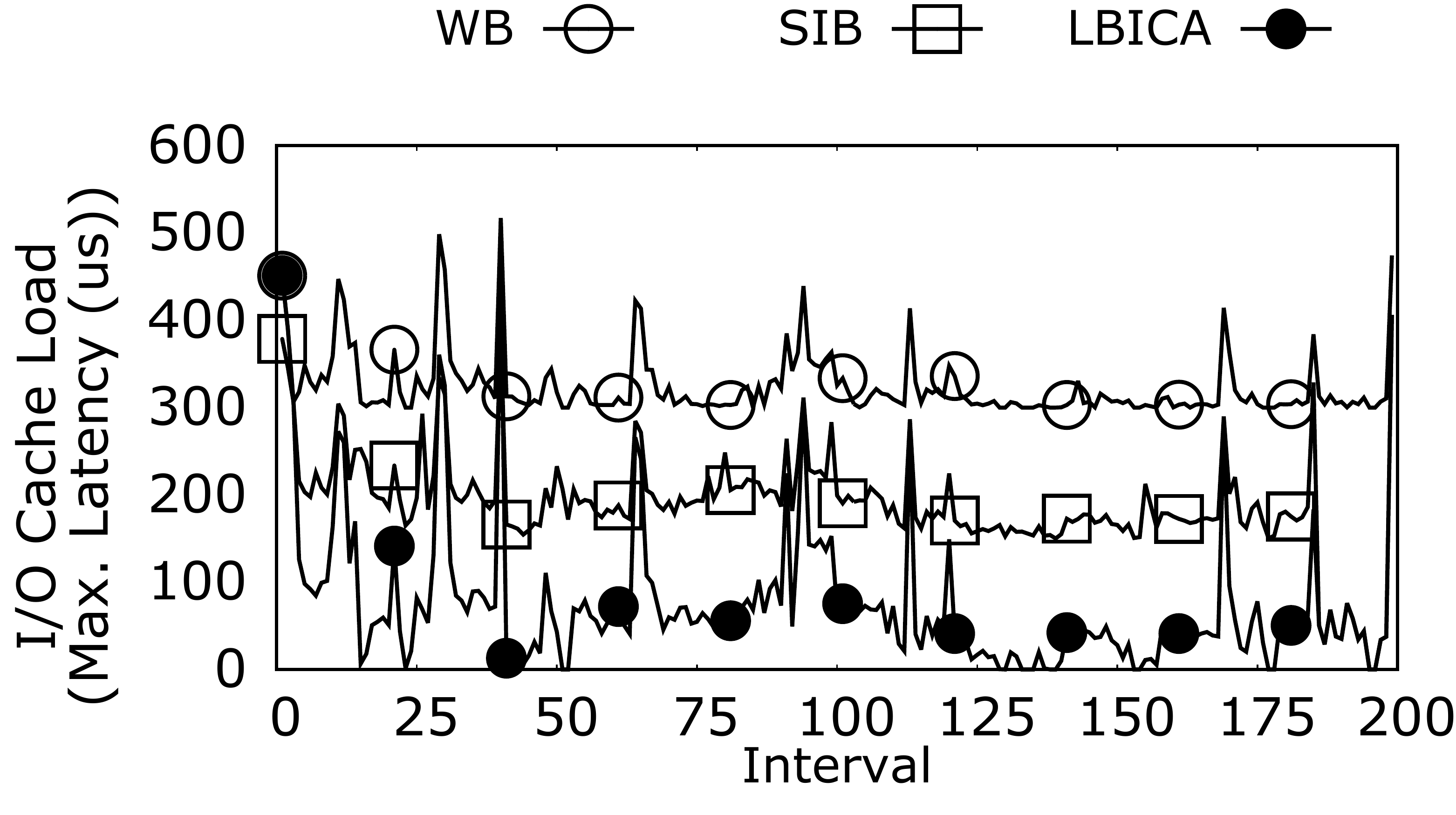}%
		\label{fig:tpcc}}
	\hfil
	\hspace{-0.4em}
	\subfloat[Mail Server]{\includegraphics[width=.31\textwidth]{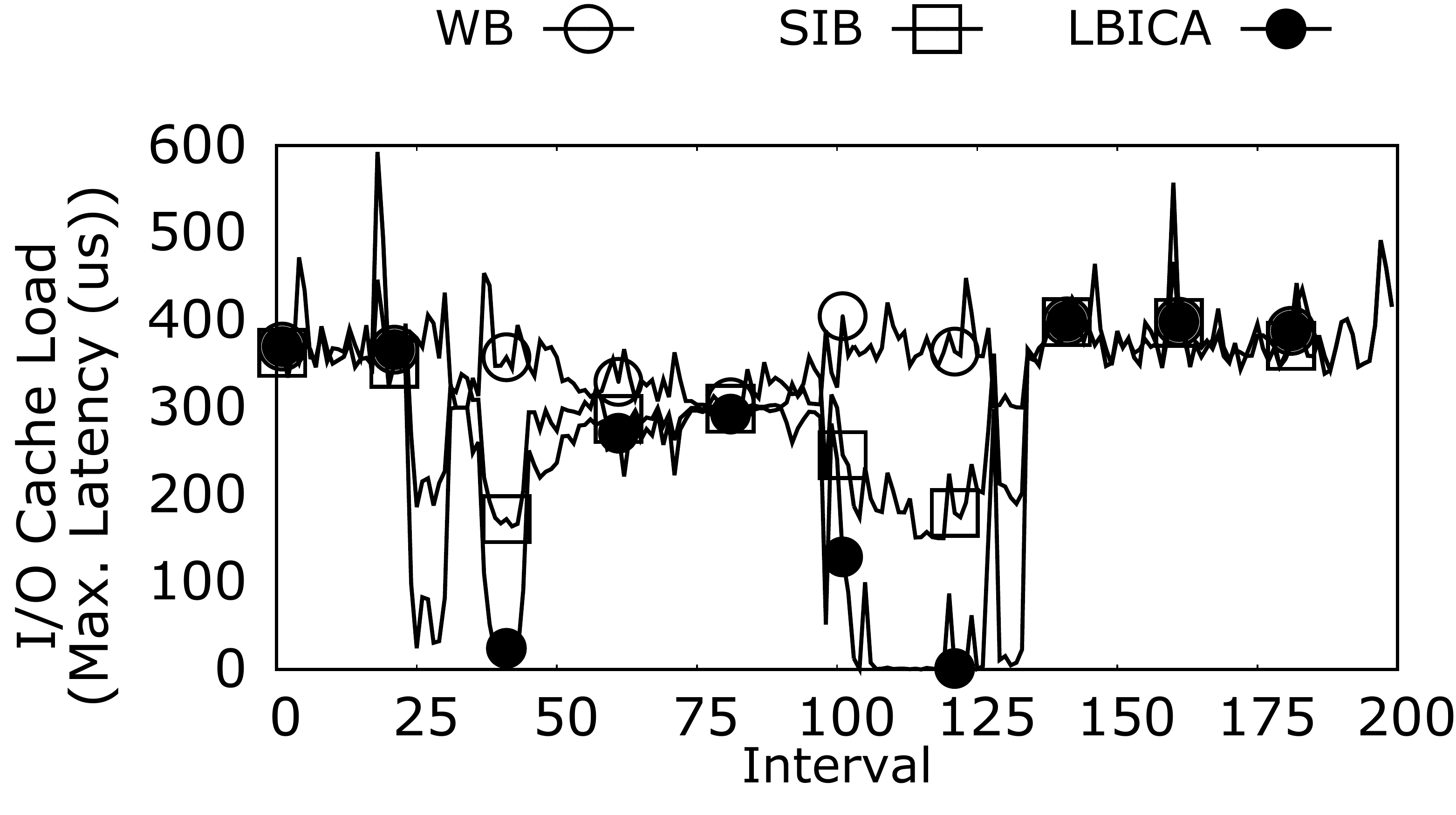}%
		\label{fig:mail}}
	\hfil
	\subfloat[Web Server]{\includegraphics[width=.31\textwidth]{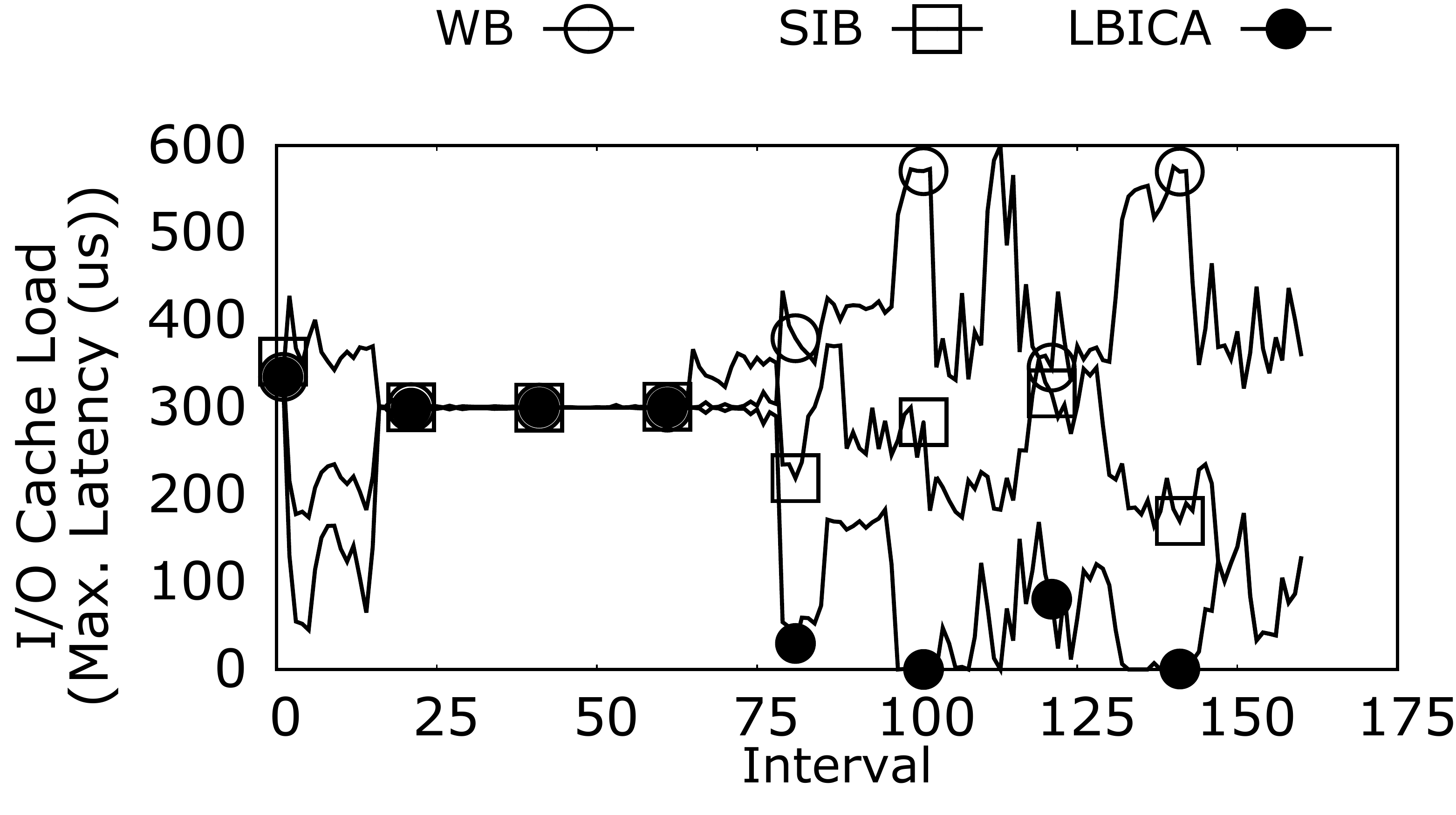}%
		\label{fig:web}}
	
		\vspace{-0.5em}
	\caption{I/O load (in terms of max. latency) on the I/O cache  by WB, SIB, and LBICA.}
		\vspace{-0.5em}
	\label{fig:load}
\end{figure*}

\begin{figure*}[!htb]
	\centering
		\vspace{-1em}
	\subfloat[TPC-C]{\includegraphics[width=.31\textwidth]{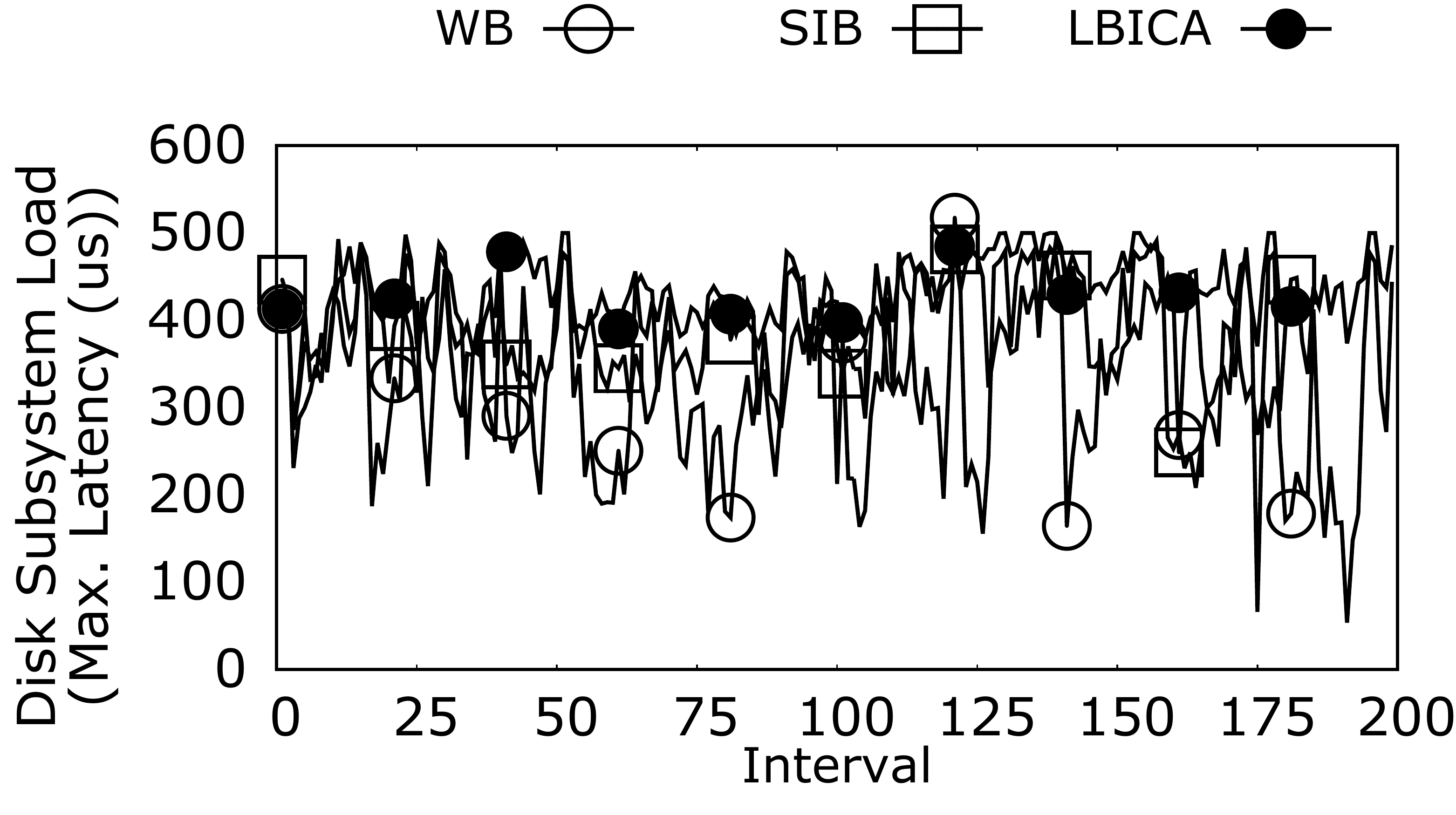}%
		\label{fig:tpcc-hdd}}
	\hfil
	\hspace{-0.4em}
	\subfloat[Mail Server]{\includegraphics[width=.31\textwidth]{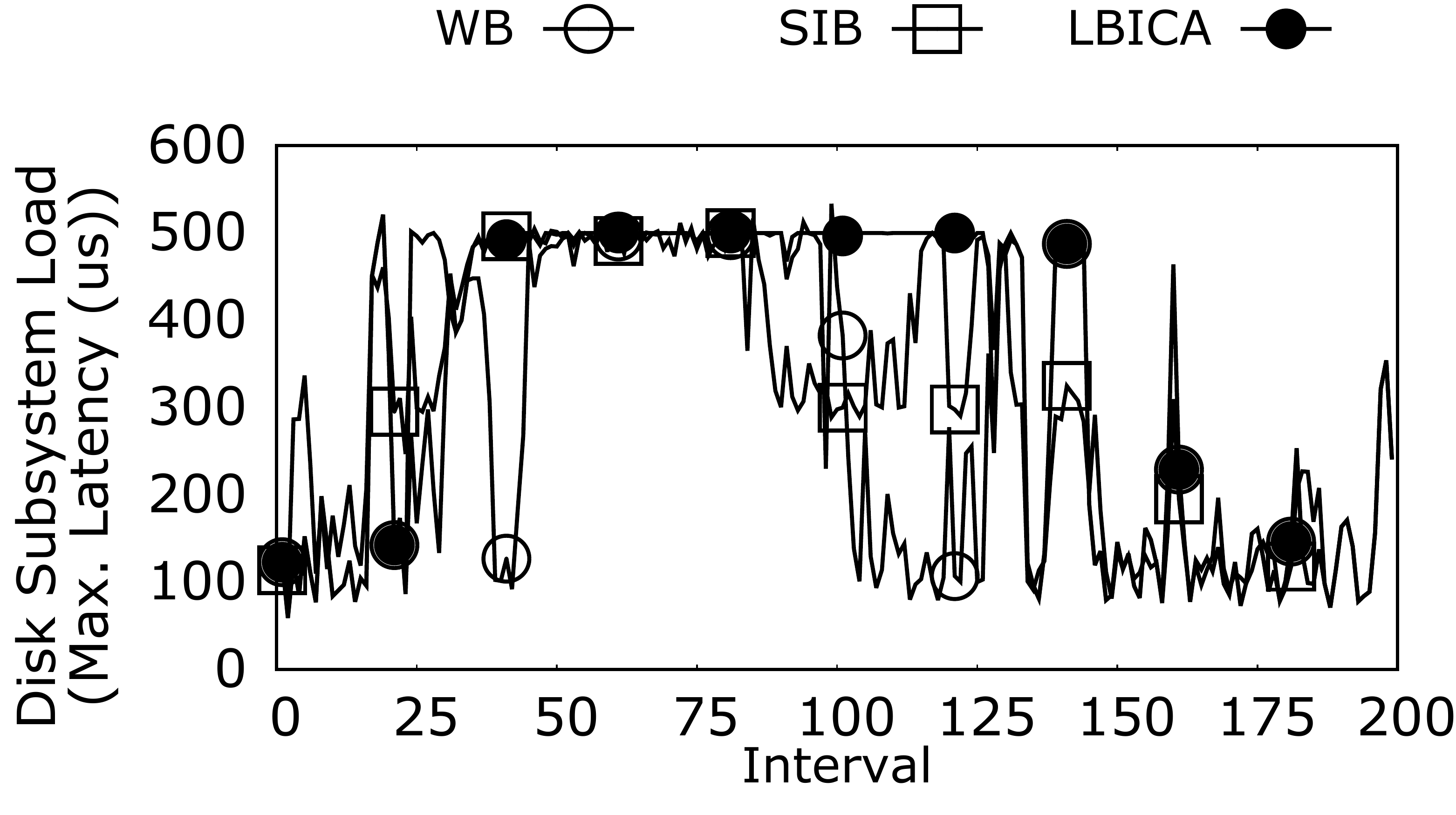}%
		\label{fig:mail-hdd}}
	\hfil
	\subfloat[Web Server]{\includegraphics[width=.31\textwidth]{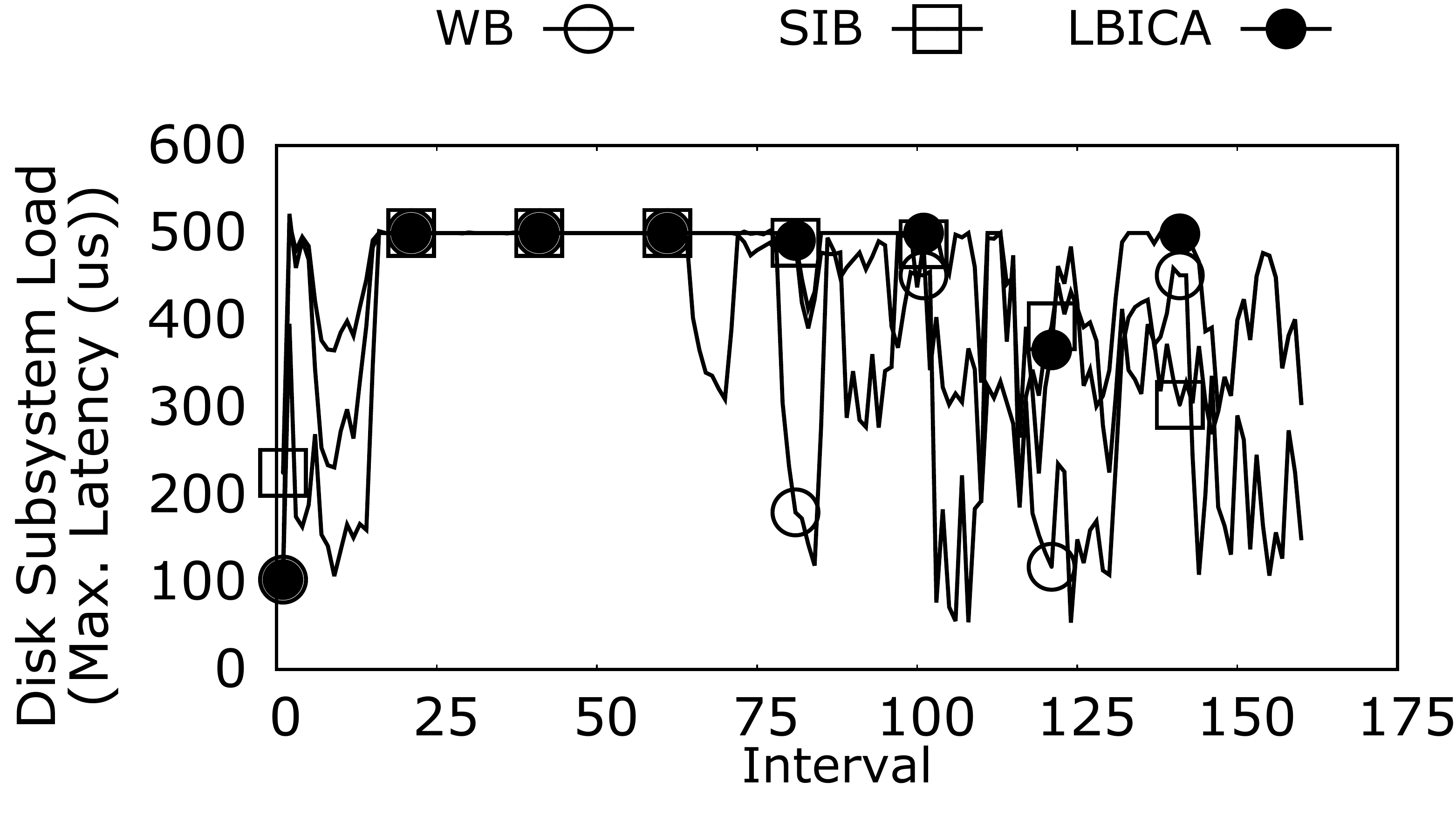}%
		\label{fig:web-hdd}}
	
		\vspace{-0.5em}
	\caption{I/O load (in terms of max. latency) on disk subsystem by WB, SIB, and LBICA.}
		\vspace{-0.5em}
	\label{fig:load-hdd}
\end{figure*}

In this section, we compare the I/O load of the cache (in terms of queue size) in different architectures (baseline WB cache, SIB, and LBICA).
Fig. \ref{fig:load} and Fig. \ref{fig:load-hdd} show the I/O load on the cache and disk subsystem for different workloads provided by WB cache, SIB, and LBICA. We monitor and report the queue time and maximum latency in intervals of 10 minutes.

We make two main observations: 
\begin{enumerate}[leftmargin=*]
	\item WB cache fails in balancing the I/O load in which the SSD cache becomes the performance bottleneck of the requests in all intervals. This is because WB policy directs all requests to the cache to achieve the maximum hit ratio. In contrast, such scheme imposes significant high I/O load on the caching layer, resulting in high I/O latency.
	
	\item LBICA, compared to SIB, reduces the load on the I/O cache by 30\%, on average. The reason behind this is that LBICA, unlike SIB, assigns an efficient write policy to the I/O cache and hence, bypasses large number of requests to the disk subsystem. The bypassed requests by LBICA to the disk subsystem are served with smaller latency than that of by I/O cache (due to the large queue time of the I/O cache).  
	
\end{enumerate}

\subsection{Workload Characterization and Policy Assignment}
\begin{figure*}[!htb]
	\centering
		\vspace{-1em}
	\subfloat[TPC-C]{\includegraphics[width=.31\textwidth]{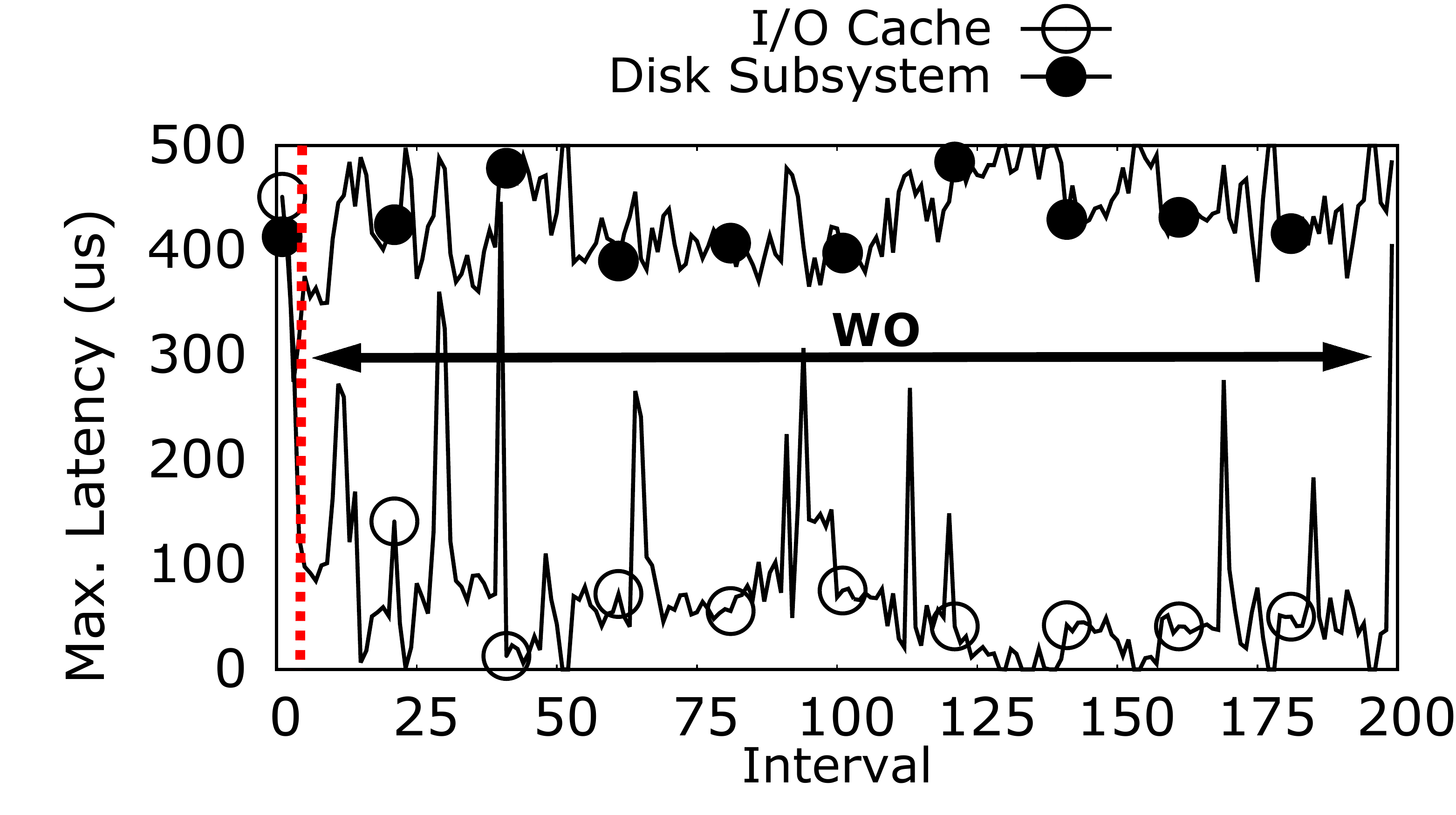}%
		\label{fig:tpcc-lbica}}
	\hfil
	\hspace{-0.4em}
	\subfloat[Mail Server]{\includegraphics[width=.31\textwidth]{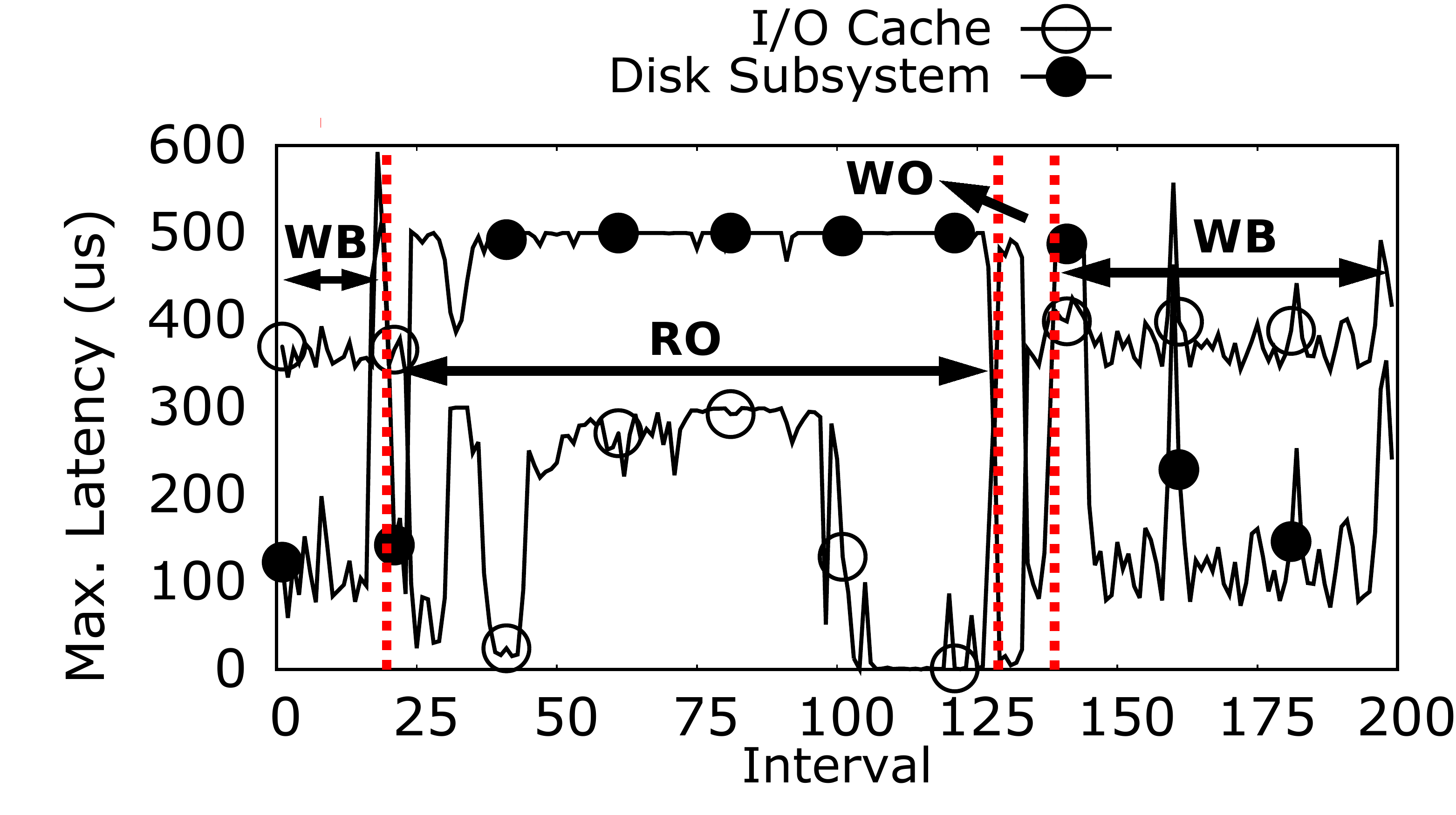}%
		\label{fig:mail-lbica}}
	\hfil
	\subfloat[Web Server]{\includegraphics[width=.31\textwidth]{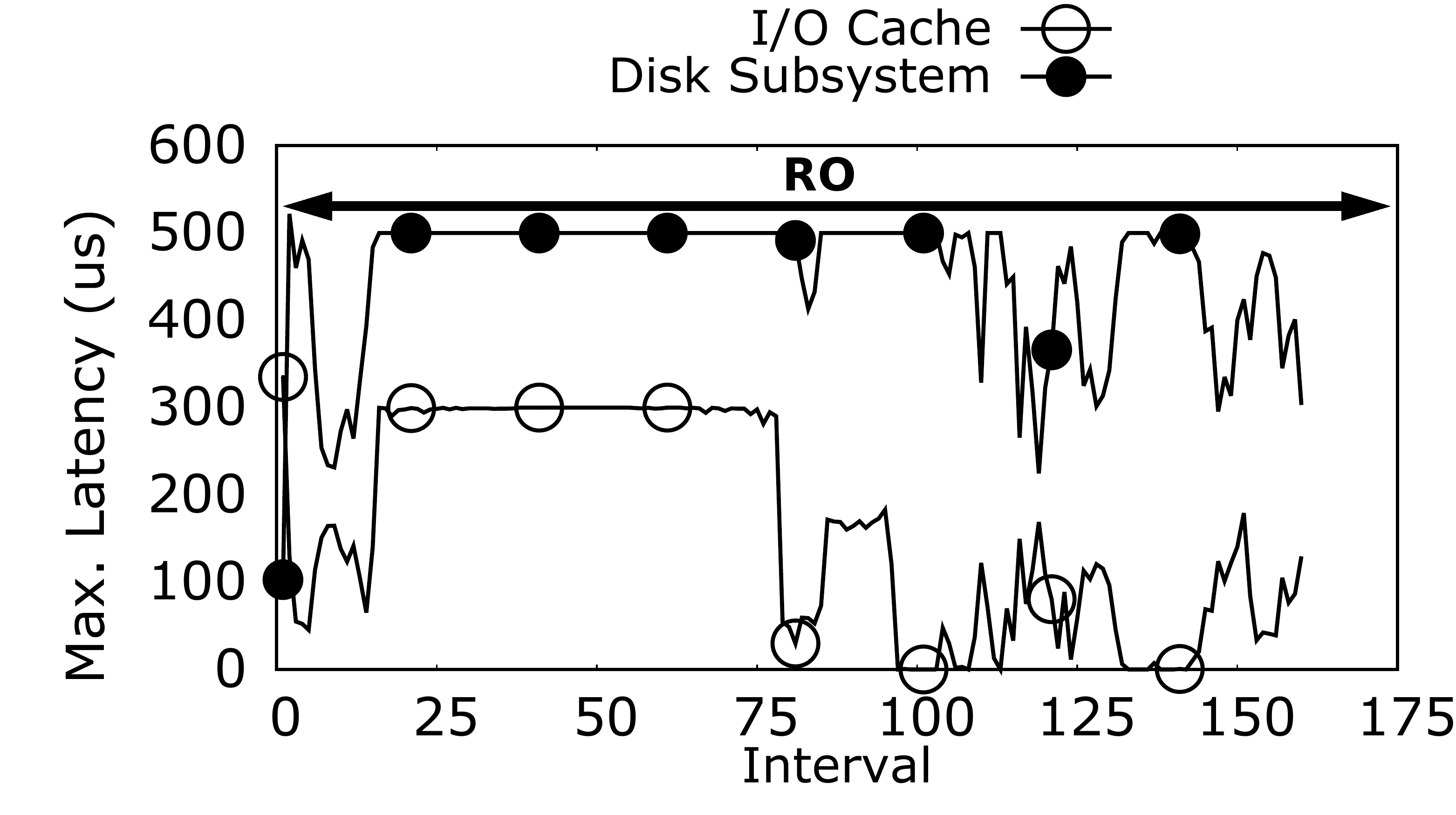}%
		\label{fig:web-lbica}}
	
		\vspace{-0.5em}
	\caption{Workload characterization and policy assignment by LBICA in burst intervals.}
		\vspace{-0.5em}
	\label{fig:load-lbica}
\end{figure*}

In this section, we show how LBICA detects the characteristics of the running burst workload and decides an efficient write policy in different intervals.
The initial write policy of the cache is set to WB, which in the burst accesses, LBICA sets different write policies (WO and RO) based on the characteristics of the running workload.
Fig. \ref{fig:load-lbica} shows the I/O load on the cache and disk subsystem by LBICA which a) the burst intervals, b) detected characteristics of the workload, and c) assigned write policy by LBICA are provided in this figure. 
We make the following observations:
\begin{enumerate}[leftmargin=*]
	\item For the TPC-C workload, at the interval of 3, LBICA reports a burst interval (in which the queue time of the I/O cache is greater than that of by disk subsystem).
	As shown in Fig. \ref{fig:tpcc-lbica}, LBICA characterizes the running workload as \emph{random read} due to the ratio of the accesses (R: 44\%, W: 2.2\%, P: 51\%, and E: 2.8\%). Thus, it assigns WO policy to the I/O cache. In this case, LBICA prevents buffering read misses on the cache, which contributes to 51\% of the load on the I/O cache (i.e., promotions (P)). Doing so, the I/O load on the SSD is reduced by more than 50\% with a negligible performance overhead.  
	
	\item For the mail server workload, at interval 23, a burst interval is detected by LBICA. Then the characterization of the workload is set to \emph{mixed read write} because of the majority of R and W operations on the I/O cache queue (R: 13.9\%, W: 70.4\%, P: 3.9\%, and E: 11.8\%). In this case, LBICA sets the RO policy on the cache, in which the read accesses are served by cache, while the write accesses are bypassed to the disk subsystem. Thus, LBICA reduces the I/O load on the cache by 70\%.
	
	Then at interval 128, a burst interval is detected, which is mainly due to R and P operations. Such workload is from \emph{random read} type which LBICA assigns WO policy on the cache. In this case, read hits and read misses are served by the cache and disk subsystem, respectively, while we prevent buffering read misses in the I/O cache, resulting in about 50\% load reduction on the I/O cache.
	
	At the rest of the experiment, at interval of 134, the I/O cache becomes the performance bottleneck. In this case, since the majority of operations in the I/O cache queue are from W and E type (about 90\% ), the workload is detected as \emph{write intensive}, and hence, LBICA assigns WB policy on the cache.  
	
	\item For the web server workload, at the first interval, LBICA detects the I/O cache as performance bottleneck. The majority of the accesses in the I/O cache queue are from R and W type (R: 17.9\%, W: 63.8\%, P: 7.9\%, and E: 10.4\%), which the workload is detected as \emph{mixed read write}. LBICA sets the RO policy on the cache, and hence, reduces 63\% of the load on the cache.
\end{enumerate} 

We conclude that LBICA, in the burst accesses, assigns an efficient write policy on the cache and reduces the I/O load on the cache up to  70\% (48\%, on average).

\subsection{Performance Improvement}
\begin{figure}[!h]
	\centering
		\vspace{-1em}
	\includegraphics[scale=.28]{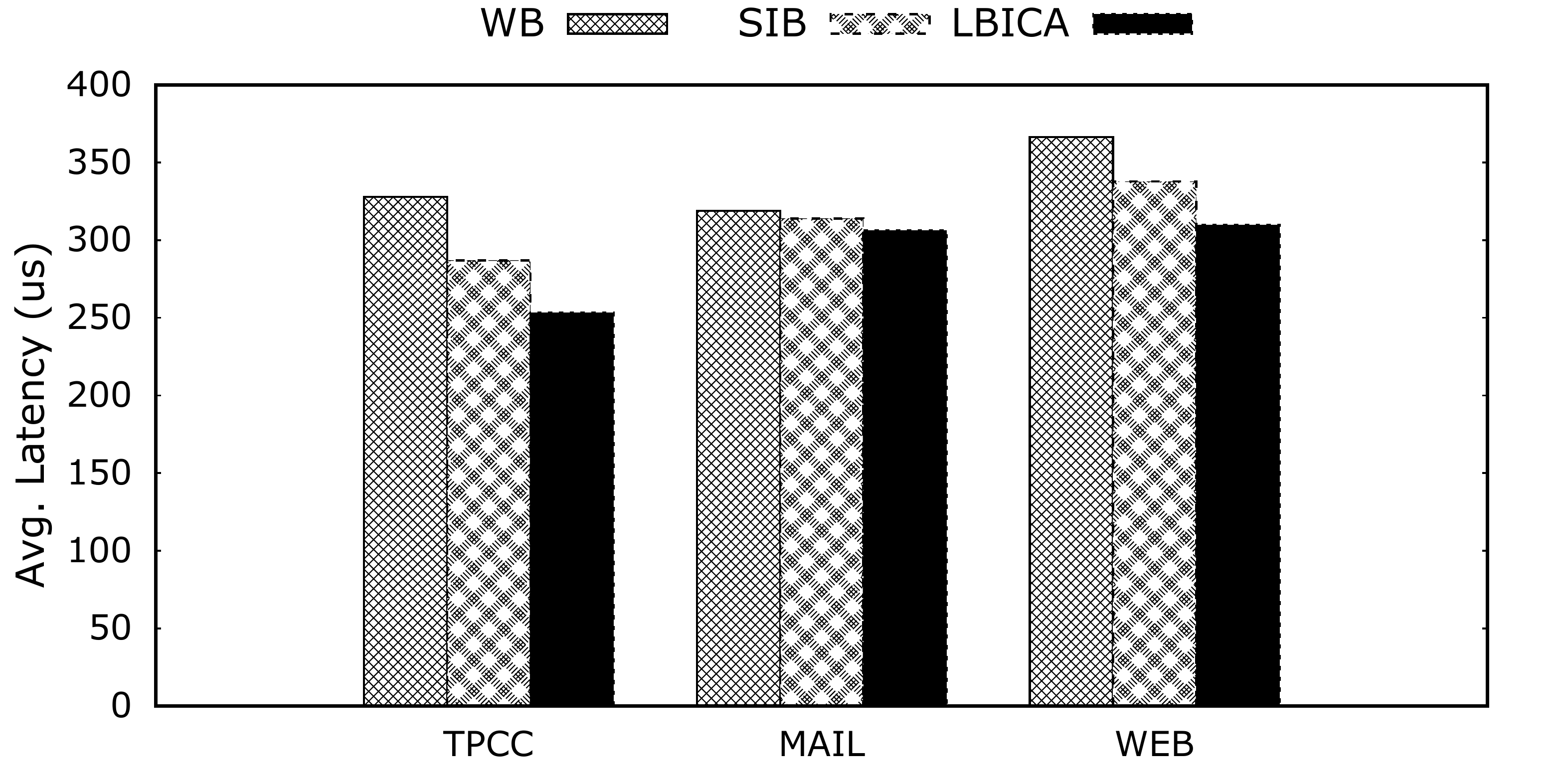}
		\vspace{-2em}
	\caption{Average latency achieved by WB cache, SIB, and LBICA.}
		\vspace{-1em}
	\label{fig:perf}
\end{figure}

In this section, we compare the average performance of the running workloads by WB cache, SIB, and LBICA.
Fig. \ref{fig:perf} shows the overall latency of the workloads during the experiments. 
We make two main observations:
\begin{enumerate}[leftmargin=*]
	\item LBICA improves the average latency up to 22\% and 11.7\% compared to WB cache and SIB, respectively (14\% and 7\%, on average).
	\item The highest performance improvement is achieved for TPC-C workload while LBICA only improves the performance of mail server by 4\%. The reason behind is that LBICA assigns RO policy in the intervals 23 to 128 and bypasses 70\% of requests (write accesses) to the disk subsystem, resulting in a poor performance improvement.
\end{enumerate}

\section{Conclusion}
\label{sec:conclusion}
In this paper, we proposed LBICA, a novel I/O cache load balancing scheme, which effectively detects burst I/O accesses and assigns efficient cache write policy to prevent the I/O cache from becoming the bottleneck. Using kernel level I/O tracing tools, LBICA analyzes the I/O load on both cache and disk subsystem. Doing so, it detects the burst intervals and characterizes the running workloads based on the type of in-queue requests. We put the residing requests in the I/O cache queue into four groups: Write (W), Read (R), Promote (P), and Evict (E). Then we characterize the running workload based on the ratio of different request types in the I/O cache queue. Based on the workload characteristics, we set an efficient write policy on the I/O cache with the purpose of balancing load on both cache and disk subsystem and providing the highest performance (in terms of latency).
Our evaluations on a physical system shows that LBICA reduces the load on the I/O cache by 48\% and improves the performance of burst accesses by 30\% compared to the latest state-of-the-art I/O cache load balancing scheme. 




%
\section*{Acknowledgment}
\vspace{-0.6em}	
	This work has been partially supported by \emph{Iran National Science Foundation} (INSF) under grant number 9606071 and by HPDS Corp.

\vspace{-0.6em}

\bibliographystyle{IEEEtran}
\bibliography{IEEEabrv,ref}

\end{document}